\documentclass{ckm}                 
\usepackage{ckm-def,url}

\usepackage{txfonts}            

\confname{Workshop on the CKM Unitarity Triangle, IPPP Durham, April 2003}

\title{WG2 Conveners' Report: $V_{td}$ and $V_{ts}$, $B$--$\bar B$
  mixing, radiative penguin and rare (semi)leptonic decays}

\author{JM Flynn\addressmark{a}, M~Paulini\addressmark{b}
and S~Willocq\addressmark{c}}

\address[a]{School of Physics \& Astronomy, University of Southampton}
\address[b]{Department of Physics, Carnegie Mellon University}
\address[c]{Department of Physics, University of Massachusetts--Amherst}

\begin{document}

\begin{abstract}
  We introduce the Working Group~2 proceedings contributions from the
  2nd workshop on the CKM Unitarity Triangle and note their connection
  to the proceedings of the first workshop.  The topic of WG\,2 was
  the determination of the CKM matrix elements $V_{td}$ and $V_{ts}$
  from $B$-$\bar B$~mixing, radiative penguin $B\rightarrow
  X_{s/d}\gamma$ decays and rare (semi)leptonic decays such as $B
  \rightarrow X_s \ell^+\ell^-$.
\end{abstract}

\maketitle


At the first CKM workshop, Working Group~2 focused on the present
status of determining the Cabibbo-Kobayashi-Maskawa matrix elements
$V_{td}$ and $V_{ts}$ from $B$~meson oscillations~\cite{yellowbook}.
On the experimental side, the status of the measurements of the
$B^0_d$ oscillation frequency $\Delta M_d$ and the limits on the
$B^0_s$~oscillation frequency $\Delta M_s$ were discussed.  From the
theoretical end, the focus was on the status and prospects of
determining the non-perturbative parameters for $B$~meson mixing from
lattice QCD calculations.

For the second CKM workshop at Durham, the experimental status of
$B$~meson mixing and the near term prospects for measuring
$B^0_S$~meson oscillations as well as the status of lattice QCD
calculations were re-evaluated. However, the focus shifted towards
methods of determining the CKM matrix elements $V_{td}$ and $V_{ts}$
from other sources in the next five years. In particular, the
experimental prospects for measuring $|V_{td}/V_{ts}|^2$ from
exclusive rare radiative decays $B\to K^* \gamma$ and $B\rightarrow
\rho\gamma$ and their theoretical limitations were explored.
Furthermore, rare (semi)leptonic decays $B \rightarrow X_s
\ell^+\ell^-$ were discussed as summarized below.

\section{$B$ Meson Mixing}

In the Standard Model, $B^0$--$\bar B^0$ mixing occurs via second order
weak processes.  The mass difference $\Delta M_q$ between the two mass
eigenstates $B_H$ and $B_L$ of the neutral $B_q$~meson ($q = d, s$)
can be determined within the Standard Model by computing the
electroweak box diagram, where the dominant contribution is through
top quark exchange:
\begin {equation}
\Delta M_q = 
\frac{G_{\rm F}^2}{6 \pi^2}\, m_{B_q}\, 
(\hat B_{B_q} F_{B_q}^2)\, \eta_B\, \mw^2 S_0(x_t)\, |V_{tq}|^2.
\label{eq:DMD}
\end{equation}
Here, $\hat B_{B_q}$ is the bag parameter of the $B$~meson, $F_{B_q}$
is the weak $B$~decay constant, $\eta_B$ is a QCD correction and
$S_0(x_t)$ is a slowly varying function of the top quark and $W$ boson
mass.

The main uncertainty in determining $V_{td}$ from $\Delta M_d$ comes
from the factor $\hat B_{B_d} F_{B_d}^2$ in Eq.~(\ref{eq:DMD}). In the
standard analysis of the Unitarity Triangle (UT), the $B^0_s$~mixing
frequency $\Delta M_s$ is used in a ratio with $\Delta M_d$ defining
the quantity $\xi$:
\[
\frac{\vtd}{|V_{ts}|}= 
  \xi\sqrt{\frac{m_{B_s}}{m_{B_d}}}
  \sqrt{\frac{\Delta M_d}{\Delta M_s}},
\qquad
\xi = 
  \frac{F_{B_s} \sqrt{\hat B_{B_s}}}{F_{B_d} \sqrt{\hat B_{B_d}}}.
\]
In the ratio the theoretical uncertainties are significantly reduced.

\subsection{Experimental Status and Prospects for $B$~Mixing}

The experimental status of $\Delta M_d$ is nicely summarized in the
contribution to the proceedings by Ronga~\cite{ronga-ckm03}. The most
recent world average value $\Delta M_d = (0.502 \pm 0.007)$~ps$^{-1}$
is dominated by results from the $e^+ e^-$~$B$~factories. It
constitutes an impressive 1.4\% measurement of the $B^0_d$~oscillation
frequency.  In the next 3--4 years, the BaBar and Belle experiments
will collect datasets of about $500\,\mathrm{fb}^{-1}$ each.  The
experimental precision on $\Delta M_d$ is expected to reduce to about
0.5\% by then.

The future interest in $B$~mixing clearly lies in the discovery of
$\Bs$~oscillations. The current limit from LEP/SLC/CDF is $\Delta M_s
> 14.4$~ps$^{-1}$ at 95\% C.L. with a combined sensitivity of
$19.3$~ps$^{-1}$. The place to observe $\Bs$~mixing in the next few
years is the Fermilab Tevatron. The contribution by
Lucchesi~\cite{lucchesi-ckm03} summarizes the prospects of the CDF and
D\O\ experiments to measure $\Delta M_s$ in Run\,II. Using data
collected with the new hadronic track trigger, CDF has reconstructed
the fully hadronic decay $B_s^0 \rightarrow D_s^- \pi^+$ observing
$44\pm11$ $B_s$~events with $D_s^- \rightarrow \phi \pi^-,\ 
\phi\rightarrow K^+K^-$. This sample is used first for initial
measurements of $B_s^0$~production fraction and branching ratios. Work
is in progress to quantify the expectations for a measurement of
$B_s^0$~mixing with the first Run\,II events now in hand.  Both CDF
and D\O\ were also able to collect initial samples of
$B_s^0$~particles from semileptonic decays or from $B_s^0 \rightarrow
J/\psi \phi$ which are used for preliminary $B_s^0$ mass and lifetime
measurements~\cite{lucchesi-ckm03}.

The prospects for measuring $\Delta M_s$ with the Atlas detector at
the LHC were presented by Ghete~\cite{ghete-ckm03}.  A detailed
treatment of systematic uncertainties was performed and the impact of
changes to the luminosity target for the LHC start-up were discussed.

\subsection{Lattice QCD Results for $\xi$}

In the report from the first CKM workshop~\cite{yellowbook}, the
lattice QCD result for $\xi$ is quoted with an asymmetric upper error
to account for the effects of chiral logarithms in extrapolating from
simulated quark masses to their real-world values. For UT analyses
this has the practical consequence of raising the central value of
$\xi$. Ultimately the extrapolation can be controlled by simulations
with lighter quarks, but in the interim, there is some profit in
investigating ways to reduce the sensitivity to chiral logarithms.  At
the Durham meeting Becirevic~\cite{beci-ckm03} highlighted one way to
accomplish this, by considering the ratio $(\fB s/\fB d)/(f_K/f_\pi)$,
where the chiral logarithms largely cancel.  However one has to assume
that the cancellation, which is shown theoretically for light quarks,
still holds in the region where one matches onto lattice results,
evaluated for heavier quark masses. The current error on $\xi$ is
$5$--$10\%$ and will certainly be reduced with results from new
simulations with lighter quarks. Experimental input from CLEOc on
$F_{D_s}/F_D$ will also help.

\section{Rare Radiative and Semileptonic $B$ Decays}

At the first CKM workshop~\cite{yellowbook}, discussion centered on
rare radiative $B$ decays and is nicely summarized by Ali and
Misiak~\cite{AliMisiakYellow} in the first workshop report. At the
Durham meeting, the discussion was widened to include the rare
semileptonic decays, addressing both CKM phenomenology and new physics
reach.  On the theory side, inclusive decays were reported on by
Hurth~\cite{hurthlunghi-ckm03}, while Buchalla~\cite{buchalla-ckm03}
and Lunghi~\cite{hurthlunghi-ckm03} covered exclusive radiative
decays.  On the experimental side, Playfer~\cite{playfer-ckm03}
provided an overview, whereas Eigen~\cite{eigen-ckm03} and
Nakao~\cite{nakao-ckm03} focused on results from BaBar and Belle,
respectively.

The inclusive decays have a theoretically clean description and
observables are dominated by partonic contributions. In the exclusive
case, the difficult problem of evaluating quark operator matrix
elements between hadronic states must be addressed. By contrast,
experimentally the inclusive modes are more challenging to measure.

\subsection{Inclusive Rare Radiative $B$ Decays}

For $B\to X_s\gamma$ the NLL QCD calculation has a charm mass
renormalization scheme dependence. The most recent
calculations~\cite{gm-bxsg,bcmu-bxsg} use the $\overline{\mathrm{MS}}$
scheme for $m_c$. With this choice for $m_c$, the branching ratio
(with a cut on photon energy) is~\cite{hurthlunghi-ckm03}
\[
\mathrm{BR}(B\to X_s\gamma) = (3.70\pm0.30) \times 10^{-4}.
\]
NNLL calculations are under study, to resolve the $m_c$
scheme-dependence issue.

Measurements of the branching ratio have been performed with several
techniques, either fully inclusive (by requiring the presence of a
high-momentum lepton in the event) or semi-inclusive (by fully
reconstructing a fraction of all possible $X_s \gamma$ final states).
Playfer~\cite{playfer-ckm03} presented a world average value of
\[
\mathrm{BR}(B\to X_s\gamma) =
 (3.47\pm0.23\pm0.32\pm0.35) \times 10^{-4},
\]
where the errors are due to statistics, experimental systematics and
theory. This world average is in good agreement with the above
theoretical prediction.

Most of the theoretical apparatus used for the $B\to X_s\gamma$
transition can be carried over to the $B\to X_d\gamma$ case, but one
should take into account the possibility of long-distance
contributions from intermediate $u$ quarks in the penguin loop.
Several arguments suggest that these long-distance effects are small.
Nonetheless, the theoretical status of $B\to X_d\gamma$ is not as
clean as $B\to X_s\gamma$.

For CKM phenomenology, $b\to s$ transitions have no relevant impact
unless unitarity is assumed, whereas $b\to d$ transitions could
provide important constraints. In the ratio $R(d\gamma/s\gamma) =
\mathrm{BR}(B\to X_d\gamma)/\mathrm{BR}(B\to X_s\gamma)$, a good part
of the theoretical uncertainty cancels, making it valuable for CKM
phenomenology, but also for new physics (since the suppression by
$|V_{td}/V_{ts}|^2$ may not hold in extended models).

For more details (including a discussion of CP asymmetries) see the
conference report by Hurth and Lunghi~\cite{hurthlunghi-ckm03} and
references therein.

\subsection{Exclusive Rare Radiative $B$ Decays}

Updates on exclusive rare radiative decays ($B\to V\gamma$ with
$V=K^*, \rho, \omega$) and their new physics impact were presented at
Durham by Buchalla~\cite{buchalla-ckm03} and
Lunghi~\cite{hurthlunghi-ckm03}.

Much interest centers on the ratios
\[
R^\pm(\rho\gamma/K^*\gamma) = 
 \left|{V_{td}\over V_{ts}}\right|^2
 {(M_B^2-M_\rho^2)^3\over(M_B^2-M_{K^*}^2)^3}
 \zeta^2 (1+\Delta R^\pm)
\]
and
\[
R^0(\rho\gamma/K^*\gamma) = {1\over2}
 \left|{V_{td}\over V_{ts}}\right|^2
 {(M_B^2-M_\rho^2)^3\over(M_B^2-M_{K^*}^2)^3}
 \zeta^2 (1+\Delta R^0)
\]
for charged and neutral decays respectively, where $\zeta =
\xi_\perp^{\,\rho}(0)/\xi_\perp^{K^*}(0)$ is a ratio of form factors
at $q^2=0$ and $\Delta R^{\pm,0}$ account for explicit
$O(\alpha_\mathrm{s})$ corrections (together with annihilation
contributions for charged $B$ decays).

$B \to K^*\gamma$ decays are well established experimentally but only
limits exist for $B \to \rho\gamma, \omega\gamma$ decays.  This is
partly due to the low expected branching ratio ($\sim 10^{-6}$) and
the difficulty in cleanly selecting $\rho$ and $\omega$ mesons.  The
best current limit on the combined charged and neutral modes is
$\mathrm{BR}(B \to \rho\gamma) < 1.9 \times 10^{-6}$ at the 90\% CL.
This translates into an experimental $90\%$ CL upper limit on the
ratio of branching ratios,
\[
R(\rho\gamma/K^*\gamma) \equiv {\mathrm{BR}(B\to\rho\gamma) \over
 \mathrm{BR}(B\to K^*\gamma)} < 0.047,
\]
which is typically a factor of $2$ above Standard Model estimates, but is
already a significant constraint on beyond-the-SM scenarios.

The form factor ratio, $\zeta$, has been estimated in a variety of
sum-rule and quark model calculations. For the workshop reports, a
value $\zeta=0.76\pm0.10$ was adopted in~\cite{hurthlunghi-ckm03} and
$1/\zeta = 1.33\pm0.13$ in~\cite{buchalla-ckm03}.  This is a
challenging quantity to evaluate in lattice QCD calculations, since
reaching $q^2=0$ necessitates a large spatial momentum for the light
meson, with the possibility of large discretization errors. One
approach is to calculate the form factors for $P\to V\gamma$ for a
range of heavy pseudoscalar meson, $P$, masses below the $B$ mass and
then extrapolate. New preliminary lattice
results~\cite{beci-ringberg03} suggest a higher value for $\zeta$
(less $\mathrm{SU}(3)$ breaking), leading to a stronger constraint on
the unitarity triangle (see also the discussions
in~\cite{hurthlunghi-ckm03} and~\cite{buchalla-ckm03}). A further
difficulty for lattice calculations will be to deal with the $\rho$
and $K^*$ in simulations where quark masses are small enough for them
to decay.

Once measurements of both charged and neutral $B\to\rho\gamma$ decays
become available, isospin-violating ratios, with numerators
proportional to $2\Gamma(B^0\to\rho^0\gamma) -
\Gamma(B^\pm\to\rho^\pm\gamma)$, could provide a useful CKM
constraint~\cite{buchalla-ckm03}.

The prospects for measuring $\mathrm{BR}(B\to\rho\gamma)$ at the $B$
Factories are good. Both BaBar and Belle have sensitivity to observe a
$5 \sigma$ signal with integrated luminosities of
$500\,\mathrm{fb}^{-1}$, as should become available in $3$ to $4$
years.  A determination of $|{V_{td}/V_{ts}}|$ should then be possible
with a precision of $\sim 15-20\%$, including theoretical and
experimental uncertainties.  A measurement of this ratio from $B$
mixing is expected to achieve higher precision but the determination
from radiative penguin decays provides a complementary approach with
different sensitivity to New Physics.

\subsection{Inclusive Rare Semileptonic Decay $B\to X_s \ell^+\ell^-$}

Since the first workshop, there has been a lot of experimental progress
in exploring the rare decays that proceed via a $b \to s \ell^+ \ell^-$
transition. The exclusive decay $B \to K\ell^+\ell^-$ has been
established by both BaBar and Belle, and the branching ratio
for the inclusive decay $B \to X_s\:\ell^+\ell^-$ has been
measured by Belle:
\[
 \mathrm{BR}(B\to X_s\:\ell^+\ell^-) =
 (6.1 \pm 1.4 ^{+1.4}_{-1.1})\times 10^{-6},
\]
for $m(\ell^+\ell^-) > 0.2$ GeV/$c^2$.
This result is in good agreement with the SM-based predictions.

The decay $B\to X_s \ell^+ \ell^-$ is dominated by perturbative corrections
once the $c\bar c$ resonances that show up as large peaks in the
dilepton invariant mass spectrum are removed. In the perturbative
`windows' outside the resonance regions, a theoretical evaluation with
a precision comparable to that in $B\to X_s\gamma$ should be
achievable, but it will be important to compare theory and experiment
using the same energy cuts to avoid any extrapolation.

The partonic calculation has now been pushed to NNLL
order~\cite{aagw,ghiy,bmu}, with the following result for the
low-$\hat s$ ($\hat s = q^2/m_b^2$) window:
\[
\mathrm{BR}(B\to X_s \ell^+\ell^-)_{\hat s \in[0.05,0.25]} =
  (1.36 \pm 0.08) \times 10^{-6}
\]
where the error is for the renormalization scale uncertainty. The
calculation includes nonperturbative contributions scaling like
$1/m_b^2$ and $1/m_c^2$. The NNLL contributions change the central
value by more than $10\%$ and significantly reduce some systematic
errors (see~\cite{hurthlunghi-ckm03} and references therein).

\subsection{Forward-Backward Charge Asymmetry in $B\to X_s \ell^+\ell^-$ and
$B\to K^* \ell^+\ell^-$}

The forward-backward asymmetry, $A_{\mathrm{FB}}(\hat s)$, uses the
angle between the $\ell^+$ and $B$ in the lepton-pair rest frame. The
position of the zero, defined by the value $\hat s_0$ for which
$A_{\mathrm{FB}}(\hat s_0)=0$, is particularly interesting since it
depends on the relative sign and magnitude of the Wilson coefficients
$C_7$ and $C_9$ and is extremely sensitive to new physics effects.

In the inclusive case, NNLL calculations~\cite{aagw,ghiy} shifted
the position of the zero and improved the precision of its location.
They also improved the prediction of the $\hat s$ (or $q^2$) shape of
the asymmetry.

For the exclusive case, the asymmetry calculation involves values for
ratios of form factors, hard-to-evaluate nonperturbative quantities.
However, the position of the zero depends only on transverse form
factors which are all related to a single function in the heavy ($b$)
quark limit, and thus is quite well-determined~\cite{bfs}.

In both inclusive and exclusive cases, discovery of a zero in
$A_{\mathrm{FB}}$ would be extremely interesting.

\section*{Acknowledgments}

We would like to thank all speakers and other participants in the
Working Group as well as the organizers of this stimulating workshop.

\bibliographystyle{ckm}
\bibliography{ckm}

\begin{thebibliography}{10}

\bibitem{yellowbook}
M.~Battaglia et~al., eds., The CKM matrix and the unitarity triangle ({CERN},
  2003), hep-ph/0304132.

\bibitem{ronga-ckm03}
F.J. Ronga, in Ball et~al.  \cite{CKMeConf} hep-ex/0306061.

\bibitem{lucchesi-ckm03}
D.~Lucchesi, in Ball et~al.  \cite{CKMeConf} hep-ex/0307025.

\bibitem{ghete-ckm03}
B.~Epp, V.M. Ghete and A.~Nairz, in Ball et~al.  \cite{CKMeConf}
  hep-ph/0307114.

\bibitem{beci-ckm03}
D.~Becirevic, in Ball et~al.  \cite{CKMeConf} hep-ph/0310072.

\bibitem{AliMisiakYellow}
A.~Ali and M.~Misiak, Radiative Rare {$B$} Decays, in Battaglia et~al.
  \cite{yellowbook} chap.~6, p. 285, hep-ph/0304132.

\bibitem{hurthlunghi-ckm03}
T.~Hurth and E.~Lunghi, in Ball et~al.  \cite{CKMeConf} hep-ph/0307142.

\bibitem{buchalla-ckm03}
S.~Bosch and G.~Buchalla, in Ball et~al.  \cite{CKMeConf} .

\bibitem{playfer-ckm03}
S.~Playfer, in Ball et~al.  \cite{CKMeConf} hep-ex/0308004.

\bibitem{eigen-ckm03}
G.~Eigen, in Ball et~al.  \cite{CKMeConf} .

\bibitem{nakao-ckm03}
M.~Nakao, in Ball et~al.  \cite{CKMeConf} hep-ph/0307031.

\bibitem{gm-bxsg}
P.~Gambino and M.~Misiak, Nucl. Phys. \textbf{B611} (2001) 338, hep-ph/0104034.

\bibitem{bcmu-bxsg}
A.J. Buras, A.~Czarnecki, M.~Misiak and J.~Urban, Nucl. Phys. \textbf{B631}
  (2002) 219, hep-ph/0203135.

\bibitem{beci-ringberg03}
D.~Becirevic, in Ringberg Phenomenology Workshop on Heavy Flavours (2003)
  \url{http://wwwth.mppmu.mpg.de/members/ahoang/ringberg2003/page07.html}.

\bibitem{aagw}
H.H. Asatrian, H.M. Asatrian, C.~Greub and M.~Walker, Phys. Rev. \textbf{D66}
  (2002) 034009, hep-ph/0204341.

\bibitem{ghiy}
A.~Ghinculov, T.~Hurth, G.~Isidori and Y.P. Yao, Nucl. Phys. \textbf{B648}
  (2003) 254, hep-ph/0208088.

\bibitem{bmu}
C.~Bobeth, M.~Misiak and J.~Urban, Nucl. Phys. \textbf{B574} (2000) 291,
  hep-ph/9910220.

\bibitem{bfs}
M.~Beneke, T.~Feldmann and D.~Seidel, Nucl. Phys. \textbf{B612} (2001) 25,
  hep-ph/0106067.

\bibitem{CKMeConf}
P.~Ball, J.M. Flynn, P.~Kluit and A.~Stocchi, eds., 2nd Workshop on the CKM
  Unitarity Triangle, IPPP Durham, April 2003 (Electronic Proceedings Archive
  {eConf} {C0304052}, 2003).

\end{thebibliography}

\end{document}